\documentclass[review]{elsarticle}

\usepackage{lineno,hyperref}
\modulolinenumbers[5]
\usepackage{hyperref}

\usepackage{graphicx}
\usepackage{amsmath}
\usepackage{amsfonts}
\usepackage{amssymb}

\begin{document}

\begin{frontmatter}

\title{Lateral instability in a discharge channel}

\author[URJC]{M. Array\'as\corref{mycorrespondingauthor}}
\author[ICMAT]{M. A. Fontelos}

\address[URJC]{\'Area de Electromagnetismo, Universidad Rey Juan Carlos, Tulip\'an s/n, 28933 M\'ostoles, Madrid, Spain.}
\address[ICMAT]{Instituto  de  Ciencias  Matem\'aticas,(ICMAT,   
  CSIC-UAM-UCM-UC3M),Campus de Cantoblanco,  28006 Madrid, Spain.}


\cortext[mycorrespondingauthor]{Corresponding author}

\begin{abstract}
In this paper we study the lateral instability in a discharge channel using a continuum model. We observe similarities to Kelvin-Helmholz instability in fluids. In strong electric fields, lateral long wave length perturbations can grow while small wave length perturbations diminish during the discharge evolution. We perform numerical simulations and carry out asymptotic analysis of the instability.
\end{abstract}

\begin{keyword}
Lateral instability\sep discharge channel \sep Kelvin-Helmholz
\end{keyword}

\end{frontmatter}


\section{Introduction}
Electric discharges pose interesting questions from both applied and fundamental research. They appear in a vast variety of phenomena, ranging from dielectric breakdown in low temperature plasmas and semiconductors \cite{Kolobov13,Kyuregyan14,Ebert96}. In the last decades progress has been made in the understanding of different aspects of the phenomena of ionization fronts such as of the study of the branching \cite{Arrayas02}. In the case of the study of branching, the problem has some analogies to a Laplacian growth instability where the a plasma channel propagates in a neutral media, driven by an external field. There is a competing effect of the drift and the diffusion of the charge that brings the growth an instability to the propagating front \cite{Arrayas05,Arrayas08,Arrayas12}. 

The situation investigated here is the following: the discharge goes in the horizontal direction triggered by the external field which points in that direction. We consider the boundary of the discharge parallel to the initial external field (unlike previous works where the ionization front is perpendicular to the external electric field). As a result, a boundary layer of charge imbalance appears and the study of its stability is the subject of the present work. Close enough we can approximate the interface as a planar one, and introduce a small geometrical perturbation of the charge distribution. We study how this perturbation grows or decreases. Note that the drift motion of the charge is parallel to the streamer surface and the perturbation of the straight path occurs transversal to the electric field as depicted in Fig.~\ref{planar}. 
 
Our results yield an instability for long wave modes at an interface which is expanding under strong field conditions at constant velocity. In this sense, it is similar to the situation arising in other context such as the development of liquid filaments at the interface of two parallel fluid flows due to the Kelvin-Helmholtz instability, leading to sprouting of filamentary structures \cite{Fuster2013}. Kelvin-Helmholtz instability occurs due to a jump of the tangential mass flux across the interface. This discontinuity induces a swirling motion and characteristic spiral patterns at the interface. In the present context, charged particles are drifted by the electric field and one can expect that across the interface due to the change in the conductivity between air and plasma, charge density develops discontinuities so that an instability somehow analogous to Kelvin-Helmholtz's arises. Here we present the first numerical and analytical studies pointing in that direction. Our results shows that at early stages, the onset of the lateral instabilities resembles a Kelvin-Helmholtz one.

The outline of the paper is the following. First we introduce a continuum model of a discharge channel. This set of equations has the minimal ingredients to describe the process while they are amenable for making analytically progress. 
Then we study travelling waves solutions for the charge densities and compute them. The stability analysis is carried out by introducing a geometrical perturbation and calculating a dispersion relation for the modes. Thus we are able to identify the stable and unstable modes. In order to test the analytical results we perform numerical simulations. As predicted we find that initial perturbations of the planar fronts off small wave lengths diminishes and eventually disappears while larger wave length perturbations increase.

\section{The model}
We will take a model similar to the standard minimal one which was successfully used in the study of branching \cite{Ebert96, Arrayas02, Arrayas05}. Ionization will be the only process taken into account for the generation of charge. However we will add a saturation constraint due to the fact that the ionized channel is depleted of neutral particles so ionization is not allowed inside the plasma.

The balance equations for the electron, positive and neutral densities, $N_{e,p}$ and $N$, will be written as 
\begin{eqnarray}
\frac{\partial N_{e}}{\partial \tau} + \nabla_{\bf R}\cdot {\bf J}_{e} = S_{e},
\label{balance-electrons} \\  
\frac{\partial N_{p}}{\partial  \tau} = S_{p},
\label{balance-positive} \\
\frac{\partial N}{\partial \tau} = S.
\label{balance-neutral}
\end{eqnarray}
Note that at small time-scales the positive ion current can be neglected as it is more than two orders of magnitude smaller than the electron current. If the neutral gas is not moving on average, there will be no net transport for the neutral particles either, so the balance equations for positive and neutral particles take the form given by \eqref{balance-positive} and \eqref{balance-neutral}. The electron current can be written as ${\bf J}_e({\bf R},\tau)=-\mu_{e} {\boldsymbol{\cal E}} N_{e} - D_{e} \nabla_{\bf R} N_{e},$ where ${\boldsymbol{\cal E}}$ is the electric field and $\mu_{e}$ and $D_{e}$ are the mobility and diffusion coefficients of the electrons. The electric field evolution is governed by the Poisson equation,
\begin{equation}
\nabla_{\bf R}\cdot{\boldsymbol{\cal E}} =
\frac{e}{\varepsilon} \, \left( N_{p} -N_{e} \right) , \label{poisson}
\end{equation}
where $\varepsilon$ is the permittivity of the gas and $e$ the absolute value of the charge of positive ions $e$ (if it is not the case a $Z$ number taking into account the charge of the ions has to be introduced which can be removed with an appropriated scaling of the variables).

Now let us introduce the constrain on the production of the charged particles. We will assume that initially we have a neutral gas of density $N_0$, so at any time we will have
\begin{equation}
  \label{eq:balance}
  N_0 = N_p+N,
\end{equation}
and from that follows, that the source terms $S_p= - S$. On the other hand, due to charge conservation $S_e=S_p$, which relates all the source terms. Considering the production of charge by ionization in Townsend's approximation \cite{Arrayas08} we get $S_e = N_e \mu_{e} |{\boldsymbol{\cal E}}| N \sigma \exp\left({- {\cal E}_i /|{\boldsymbol{\cal E}}|}\right)$.

The densities can be scaled by the initial neutral gas density, thus $n_e = N_e/N_0$, $n_p = N_p/N_0$. The scattering cross section of the ionization process sets a characteristic length $R_0=1/N_0\sigma$, and the collision time $\tau_0=\varepsilon/e N_0\mu_e$ a time scale. Introducing the dimensionless coordinates ${\bf r}={\bf R}/R_0$, $t=\tau/\tau_0$, the dimensionless electric field ${\bf{E}}={\boldsymbol {\cal E}}\varepsilon \sigma/e$, and the dimensionless diffusion constant $D=D_{e}N_0\sigma^2\varepsilon/e \mu_e$, the model reads
\begin{eqnarray}
\frac{\partial n_{e}}{\partial t} &=&\nabla \cdot
\left(n_{e}{\bf{E}}+D_{e}\,\nabla n_{e}\right)
+n_{e}(1-n_p)|{\bf{E}}| e^{-\alpha/|{\bf{E}}|},
\nonumber\label{electron} \\
\frac{\partial n_{p}}{\partial t} &=&n_{e}(1-n_p)|{\bf{E}}|e^{-\alpha/|{\bf{E}}|},\nonumber \label{ion} \\
\nabla \cdot {\bf{E}}&=&n_{p}-n_{e}  \label{gauss}.
\end{eqnarray}
where $\alpha={\cal E}_i\varepsilon\sigma/e$ is a dimensionless parameter.

\begin{figure}
  \begin{center}
 \includegraphics[width=0.85\linewidth]{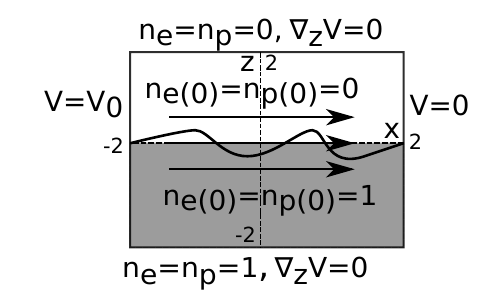}%
 \caption{A schematic of the case considered, where a planar interface parallel to the external field separates two ionized and non-ionized regions. The arrows indicate the direction of the imposed external electric field. Boundary conditions and initial conditions are also indicated and also the kind of perturbation introduced. The size of the simulation box in dimensionless units are also plotted.}
 \label{planar}
 \end{center}
\end{figure}

Close enough to the discharge channel we will assume that the interface can be approximated by a plane, so we will study the case of a expanding channel with a planar symmetry.
We will take Cartesian coordinates on the interface along the $x$-axis and $z$-axis, and a constant external field parallel to the interface, given by ${\bf E}_{ext} = E_{x0} {\bf u}_{x}$ where $E_{x0} > 0$ and ${\bf u}_{x}$ is an unitary vector in the $x$ direction, as schematically depicted in the Fig.~\ref{planar}.
We need to provide the boundary conditions for the system \eqref{gauss} according to the situation considered. Asymptotically at $z\to -\infty$ we have the discharge channel fully ionized, so $n_e=n_p=1$ and the electric field will be the external one ${\bf E} = E_{x0} {\bf u}_{x}$. At $z\to +\infty$ the system is neutral and there is not charge, so $n_e=n_p=0$. For the $x$ direction we will assume periodic conditions.

\section{Analysis of the instability}
Here we make analytical calculations for the onset of the instability observed. We expand the densities and the electric field in a small $\epsilon$ parameter in the following form 
\begin{eqnarray}
  n_e&=&\phi_e(\xi) + \epsilon n_e^{(1)}(x,z,t)+O(\epsilon^2),\nonumber\\
  n_p&=&\phi_p(\xi) + \epsilon n_p^{(1)}(x,z,t)+O(\epsilon^2),\label{ansazt}\\
  {\bf E} &=& E_{x}^{(0)}{\bf e}_x+\sqrt{D} E_z^{(0)}(\xi) {\bf e}_z+\epsilon {\bf E}^{(1)}(x,z,t)+O(\epsilon^2)\nonumber.
\end{eqnarray}
where
\begin{equation}
  \xi = \frac{z - \epsilon f(x,t)}{\sqrt{D}}-ct.
  \label{xi}
\end{equation}
Zero order terms represent travelling waves solutions and the function $f(x,t)$ represents the geometrical perturbation of such solutions. Together with the first order terms allow the asymptotic analysis of the instability.

To keep amenable the mathematical analysis and highlight the physical results, we will restrict it to the condition that $|{\bf{E}}|e^{-\alpha/|{\bf{E}}|}\approx 1$, which is a sensible choice provided that $|{\bf{E}}|\approx E_{x0}=1$, i.e. we have a strong external electric field, and the ionization critical field is lower than this external field, which allows to take the $\alpha \to 0$ limit. Under that limit, the equations \eqref{electron} get a simpler form,
\begin{eqnarray}
\frac{\partial n_{e}}{\partial t}&=&\nabla \cdot \left( n_{e} {\bf
E} + D\,\nabla n_{e} \right)+ n_{e}(1-n_p),
\label{ec1x} \\
\frac{\partial n_{p}}{\partial t} &=& n_{e}(1-n_p),  \label{ec2x} \\
\nabla \cdot {\bf E} &=& n_{p} - n_{e}.  \label{ec3x}
\end{eqnarray}

Introducing the perturbation in the form given by \eqref{ansazt} and expanding in powers of $\epsilon$, we get travelling waves in the zero order approximation.

\begin{figure}
  \begin{center}
    \includegraphics[width=0.85\linewidth]{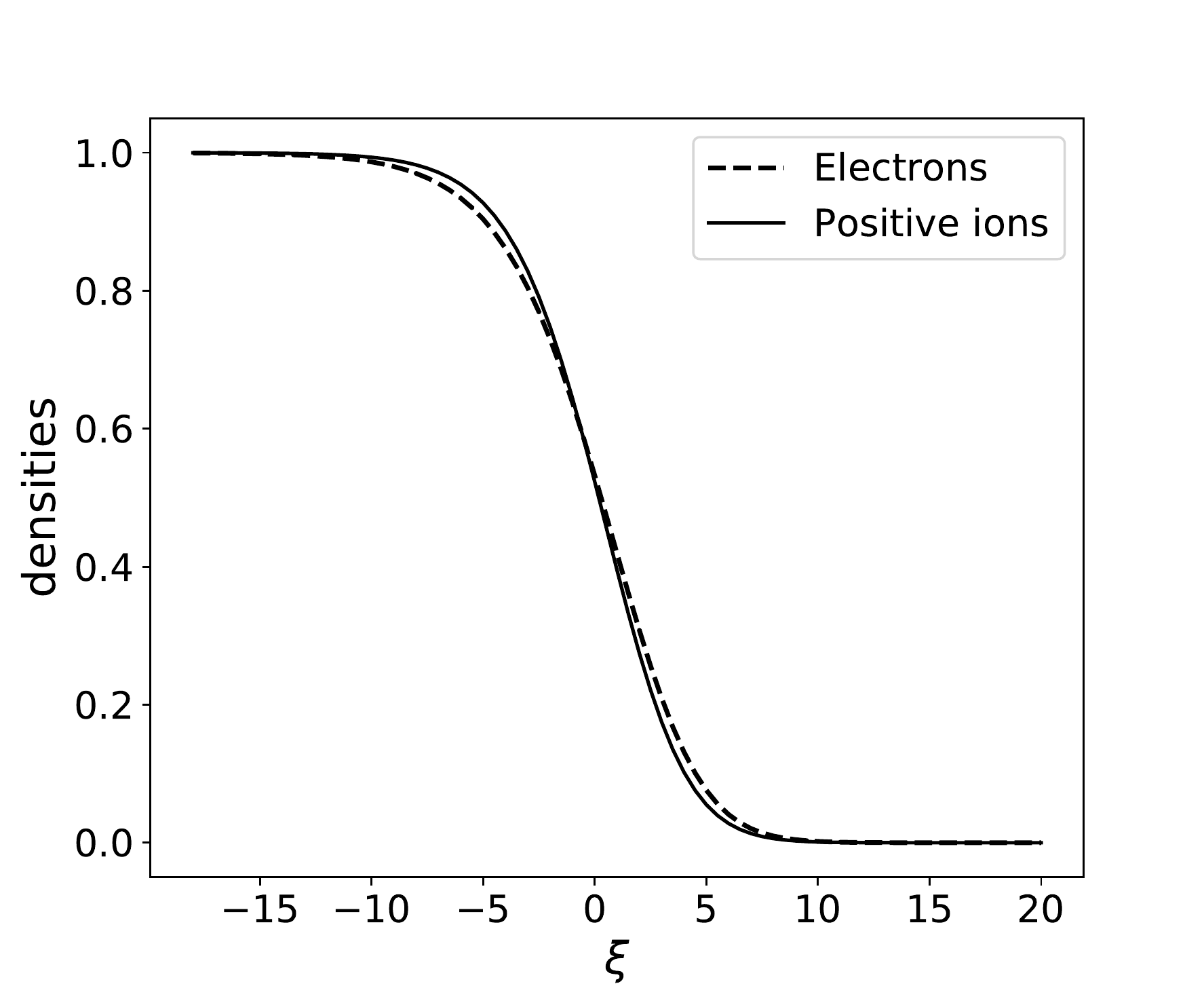}
 \caption{Negative and positive charge densities $\phi_e$ and $\phi_p$ for a planar travelling wave.}
 \label{travel1}
 \end{center}
 \end{figure}

\paragraph{Travelling waves}
The travelling waves to zero order can be obtained numerically by a shooting technique as seen in Fig.~\ref{travel1}.

We find travelling waves solutions which decay at $\xi \to -\infty$ as
\begin{eqnarray}
  \phi_e &=& 1 - A_{1} e^{\alpha_1 \xi}- A_3 e^{\beta_1\xi} \label{bcm1}\\
  \phi_p &=& 1 - B_{1} e^{\beta_1\xi}, \label{bcm2}\\
  E_z^{(0)}&=&E_3e^{\alpha_1 \xi}+E_{1} e^{\beta_1\xi}, \label{bcm3}  
\end{eqnarray}
where $A_1, A_3, B_1, E_1$ and $E_3$ are constants determined by the boundary conditions and $\alpha_1 >\beta_1$ the corresponding positive eigenvalues that can be written in terms of the wave velocity $c$. The behaviour of such travelling waves for $\xi \to +\infty$ is
\begin{eqnarray}
  \phi_e &=& A_{2} e^{-\beta_2 \xi}, \label{bcp1}\\
  \phi_p &=& B_{2} e^{-\beta_2 \xi}, \label{bcp2}\\
  E_z^{(0)}&=&E_{0}+E_{2} e^{-\beta_2 \xi}, \label{bcp3}  
\end{eqnarray}
where again $A_2, B_2, E_0$ and $E_2$ are constants and $\beta_2>0$ the corresponding eigenvalue that can be expressed as a function of c. A standard shooting technique allows then to determine the appropriate value of c so that the exponential decays in \eqref{bcm1}-\eqref{bcp3} are satisfied. Notice that to zero order, if $E_0$ is non vanishing, there is an electric field at infinity in the $z$-direction due to the contribution of the travelling wave unbalanced charged. Indeed considering the limit $D\to0$ it can be shown that the travelling wave propagating speed $c$ reaches an extremum if $E_0=0$, and in this case, $c \approx 2.0$.

\begin{figure}
  \begin{center}
  \includegraphics[width=0.8\linewidth]{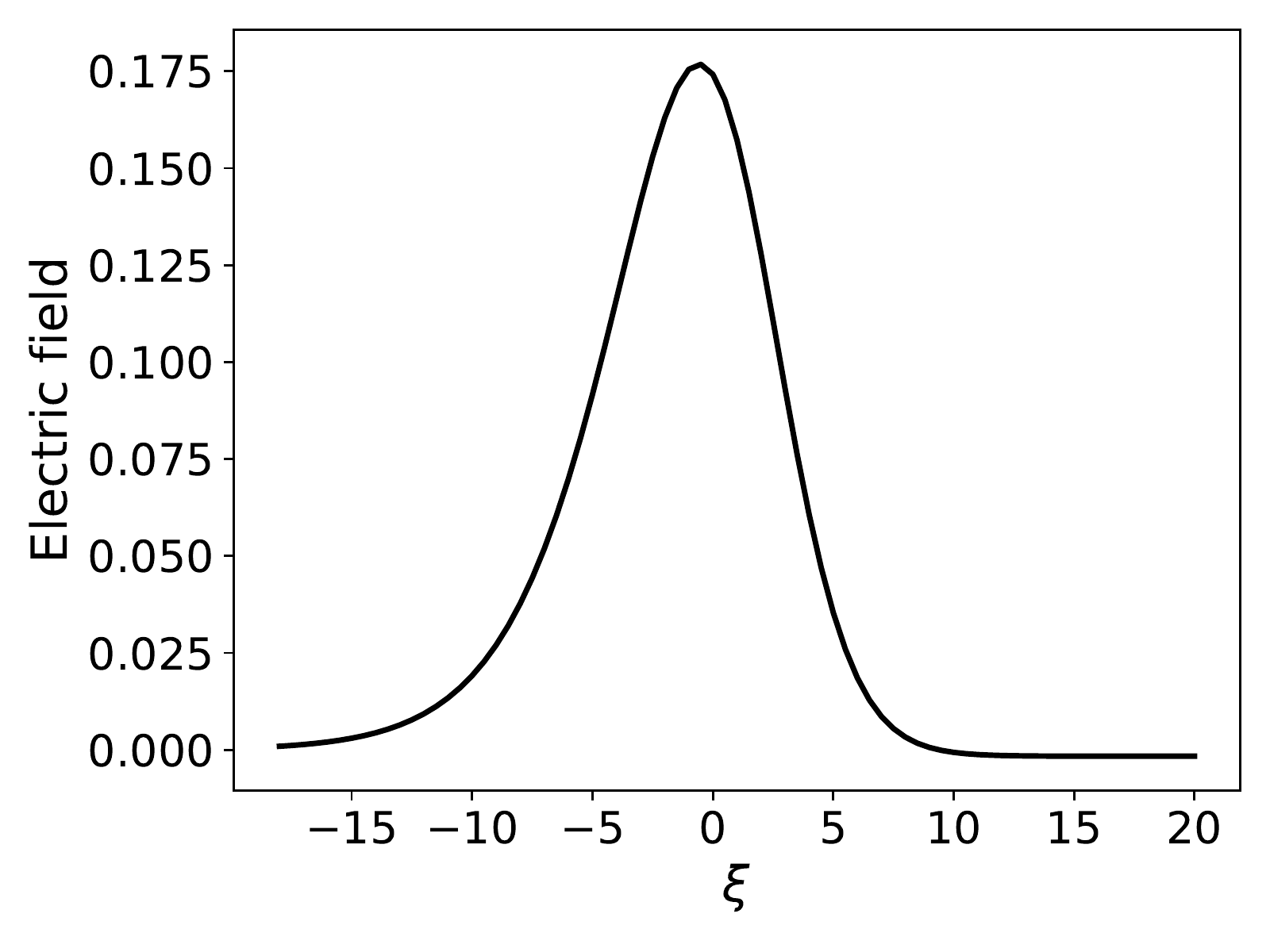}
 \caption{Electric field to zero order $E_z^{(0)}$, for the travelling wave of Fig.~\ref{travel1}, diminishing at large distances.}
 \label{travel2}
 \end{center}
 \end{figure}

The travelling wave solution for $E_0=0$ is such that the profile for negative charges is, due to their larger mobility, slightly ahead/behind the profile for positive charges as z approaches plus/minus infinity, as  shown in Fig.~\ref{travel1}. This leads to a profile for the net charge that changes sign at the interface and decays fast at infinity. Thus, charges closed to the interface in a region of $O(\sqrt{D})$ thickness are unbalanced and creates an electric field in the travelling wave direction, as it can be seen in Fig.~\ref{travel2}. Nevertheless, the fact that negative charges are mobile (due to drift by the electric field and diffusion) while the positive charges are not indicates that such dipolar configuration cannot remain stable when the interface is not flat. This fact will be visible at the next order.

\paragraph{Onset of instabilities}
Next we make a coordinate transformation $(t,x,z)\to (t,x,\xi)$ with $\xi$ given by \eqref{xi}, so \eqref{ec1x} can be arranged to read up to second order in $\epsilon$
\begin{equation}
\begin{split}
    \frac{\partial n_e}{\partial{t}}-\left(c+\epsilon\frac{f_t}{\sqrt{D}}\right)\frac{\partial n_e}{\partial\xi}-{\bf E}\cdot \left(\frac{\partial}{\partial{x}}- \epsilon\frac{f_x}{\sqrt{D}}\frac{\partial}{\partial\xi},\frac{1}{\sqrt{D}}\frac{\partial}{\partial \xi}\right)n_e&\\-D\left(\frac{\partial^2}{\partial{x}^2}+\frac{1}{D}\frac{\partial^2}{\partial \xi^2}-\epsilon\frac{f_{xx}}{\sqrt{D}}\frac{\partial^2}{\partial{x} \partial \xi} \right)= n_e(1-n_e) +O(\epsilon^2)
\end{split}
\label{travpert1}
\end{equation}

Now we have the freedom to take $f$ such it satisfies to first order, 
\begin{equation}
-f_{t}+f_{x}-E_{z}^{(1)}|_{\xi=0}+Df_{xx}=0,  \label{b1}
\end{equation}%
in which appears the $z$ component of the electric field perturbation evaluated at $\xi=0$. Choosing $f$ to fulfill \eqref{b1}, the electron density charge perturbation is governed 
\begin{equation}
\begin{split}
\frac{\partial n^{(1)}_e}{\partial{t}}-\left(c+E^{(0)}_z\right)\frac{\partial n^{(1)}_e}{\partial \xi} &- \frac{\partial n^{(1)}_e}{\partial{x}} -\epsilon\frac{\partial^2n^{(1)}_e}{\partial \xi^2}-\epsilon D \frac{\partial^2n^{(1)}_e}{\partial {{x}}^2} \\= (1-2\phi_e)n^{(1)}_e,
\label{ne1}
\end{split}
\end{equation}
after making the expansion up to $\epsilon$ order. The equation for the evolution of $n_e^{(1)}$ turns out to be a homogeneous one, so as particular solution is given by $n_e^{(1)}=0$. Electrons are mobile and may follow a geometrical perturbation without changes in their density. However, this is not the case for $n_p^{(1)}$. The physical argument is that the positive charge do not move, so the perturbation of the electron density implies the appearance of an extra charge in addition to the one created by the travelling wave, thus implies a non-trivial correction to the positive charge. Let us find this correction.

The equation \eqref{ec2x} for $n_{p}$ can be rewritten as
\[
-\frac{\partial }{\partial t}\log (1-n_{p})=n_{e},
\]%
and then, expanding in $f$ we obtain at leading order after taking the particular solution $n^{(1)}_e=0$,
\[
\left( \frac{\partial }{\partial t}-c\frac{\partial }{\partial \xi
}\right) \left( \frac{\phi _{p}^{\prime }f/\sqrt{D}+n_{p}^{(1)}}{%
1-\phi _{p}}\right) =\frac{1}{\sqrt{D}}\phi _{e}^{\prime }f.
\]

Using now the fact that for zero order travelling waves
\[
\left( -c\frac{\partial }{\partial \xi }\right) \left( \frac{%
\phi _{p}^{\prime }}{1-\phi _{p}}\right) =\phi _{e}^{\prime },
\]%
we deduce the following equation for $n^{(1)}_p$,
\begin{equation}
\frac{\partial }{\partial t}\left( \phi _{p}^{\prime }f/\sqrt{D}%
+n_{p}^{(1)}\right) -c(1-\phi _{p})\frac{\partial }{\partial
\xi }\left( \frac{n_{p}^{(1)}}{1-\phi _{p}}\right) =0.
\end{equation}
If we try now a solution of the form $n_{p}^{(1)}=N_{p}^{(1)}(\xi
)e^{\lambda t}F(x)$ and $f(x,t)=e^{\lambda t}F(x)$ we obtain
\begin{equation}
\lambda \left( \phi _{p}^{\prime }/\sqrt{D}+N_{p}^{(1)}\right) -c(1-\phi _{p})\frac{\partial }{\partial \xi }\left( \frac{N_{p}^{(1)}}{%
    1-\phi _{p}}\right) =0.
\label{Np1}
\end{equation}

We will need also the first order contribution to the electric field. Introducing a scalar electric potential 
\[
V = -x +DV^{(0)}(\xi)+\epsilon V^{(1)},
\]
we can write \eqref{ec3x} as
\[\begin{split}
-\nabla V = -\left(-1,\sqrt{D}V'^{(0)} \right)\\+
\epsilon\left(\sqrt{D}f_x V'^{(0)} +\frac{\partial{V^{(1)}}}{\partial{x}},
-\frac{1}{\sqrt{D}}\frac{\partial{V^{(1)}}}{\partial{\xi}}\right).
\end{split}\]

Thus the Laplacian to first order reads
\[
-\Delta V = -V''^{(0)} + \epsilon \sqrt{D}f_{xx}V'^{(0)} -\epsilon \Delta V^{(1)},  
\]
where $V'^{(0)}=\frac{\partial{V^{(0)}}}{\partial{\xi}}$, thus
at first order in $\epsilon$ we get expression 
\begin{equation}
  -\Delta V^{(1)}=-\sqrt{D}f_{xx}\frac{\partial V^{(0)}}{\partial \xi }+n_{p}^{(1)},
\label{potential}  
\end{equation}
This allows us to compute the electric field%
\[
E_{z}^{(1)}=-\frac{\partial V^{(1)}}{\partial z},
\]%
whose value at $\xi =0$ has to be inserted into the equation \eqref{b1} for the evolution of $f(x,t)$.

From \eqref{Np1}, using the asymptotic behaviour of the travelling waves given by \eqref{bcm2} and \eqref{bcp2}, we get the asymptotic form  of $n_{p}^{(1)}=N_{p}^{(1)}(\xi)e^{\lambda t}F(x)$  
\begin{equation}
  N_{p}^{(1)}\sim \begin{cases}
    N_{1}^{-}e^{\mu_{1}\xi },&\ \ \text{as }\xi \rightarrow -\infty,  \\
    N_{1}^{+}e^{-\mu_{2}\xi },&\ \ \text{as }\xi \rightarrow +\infty
  \end{cases}
  \label{a1}
\end{equation}
with
\begin{equation}
N_{1}^{-} =B_{1}\mu _{1}/\sqrt{D},\,\,\,
N_{1}^{+} =\frac{\lambda B_{2}\mu_{2}}{\lambda\sqrt{D} +c\mu_{2}\sqrt{D}}  \label{a4}
\end{equation}

Taking $F(x)=\exp(ikx)$ and Fourier transforms, the solution turns out to be 
\begin{eqnarray}
  {E}_{z}^{(1)} &=&-\sqrt{D}k^2{H}(\sqrt{D}k)-{G}_{1}(\sqrt{D}k)\nonumber\\
            &+&\frac{\lambda }{\sqrt{D}\lambda +\sqrt{D}c\mu _{2}}
  {G}_{2}(\sqrt{D}k),
  \label{Ez1}
\end{eqnarray}
where
\begin{equation}
{H}(s)=\frac{1}{2\pi }\int_{-\infty }^{\infty }\frac{\left( \nu
^{2}(E_{2}-E_{1})+\beta_{1}^{2}E_{2}-\beta_{2}^{2}E_{1}\right) }{
\left( s^{2}+\nu ^{2}\right) \left( \nu ^{2}+\beta_{1}^{2}\right) \left(
\nu ^{2}+\beta_{2}^{2}\right) }\nu ^{2}d\nu,
\label{Fs}
\end{equation}
and
\begin{equation}
{G}_{1,2}(s)=\frac{1}{2\pi }\int_{-\infty }^{\infty }\frac{B_{1,2}\mu _{1,2}}{
\left(s^{2}+\nu ^{2}\right) \left( \nu ^{2}+\mu _{1,2}^{2}\right) }\nu
^{2}d\nu.
\label{Gs}
\end{equation}

Thus we can use (\ref{b1}) to deduce the dispersion relation
\begin{equation}
  \begin{split}
  -\lambda+ i k-D k^{2}+Dk^{2}{H}(\sqrt{D}k)+{G}_{1}(\sqrt{D}k)\\-\frac{\lambda}{\sqrt{D}(\lambda
    +c\mu _{2})}{G}_{2}(\sqrt{D}k)=0.
  \end{split}
\end{equation}
In the limit $D$ small we can approximate it as
\begin{equation}
  \lambda\left(1+\frac{{G}_2(\sqrt{D}k)}{\sqrt{D}c\mu_2}\right)\approx i k-D k^{2}+Dk^{2}{H}(\sqrt{D}k)\nonumber+{G}_{1}(\sqrt{D}k).
  \label{apprxdisp}
 \end{equation}
The onset on the instability starts when $Re({\lambda})\to 0$, which in terms of $s_c=\sqrt{D}k_c$ implies
\begin{equation}
  {G}_{1}(s_{c})=s_{c}^{2}(1-{H}(s_{c}))
  \label{critical}
\end{equation}

One can compute travelling waves solutions as plotted in Fig.\ref{travel1} and \ref{travel2} and get the values
$\beta_{1}\simeq 0.37,\ E_{1}\simeq 0.67,\ \beta _{2}\simeq 0.8, E_{2}\simeq 2.71, \mu _{1}\simeq 0.5,\ B_{1}\simeq 0.4,\ \mu _{2}\simeq 0.9,\ B_{2}\simeq 7.4.$. Using the definitions \eqref{Fs} and \eqref{Gs} it can be numerically estimated the value of $s_{c}$ as $s_{c}\simeq 0.5$ and hence instabilities may take place for%
\begin{equation}
\sqrt{D}\frac{2\pi }{\lambda }\lesssim 0.5
\end{equation}
implying $\lambda \gtrsim 12.6\sqrt{D}$, i.e. long wave length instabilities for such values.

 \section{Numerical simulations}
 For testing the behaviour of the stability of the discharge channel, we perform some numerical simulations. We take Cartesian coordinates and put the interface at $z=0$ as sketched in Fig.~\ref{planar}.

The boundary conditions for the densities are periodic in the $x$-direction and of Dirichlet type in the $z$-direction. For the Poisson equation, zero flux is taken at the $z$ boundaries, i.e. ${\bf u}_z\cdot{\bf E}=0$ and we fix the electric potential at the $x$ boundaries so there is an external electric field of $E_{x0}=1$ intensity in dimensionless units. For the initial conditions, the interface is situated at $z=0$, with $z<0$ representing the ionized channel filled with $n_e=n_p=1$ and $z>0$ the non-ionized region with $n_e=n_p=0$.

\begin{figure}
  \begin{center}
  \includegraphics[width=0.8\textwidth]{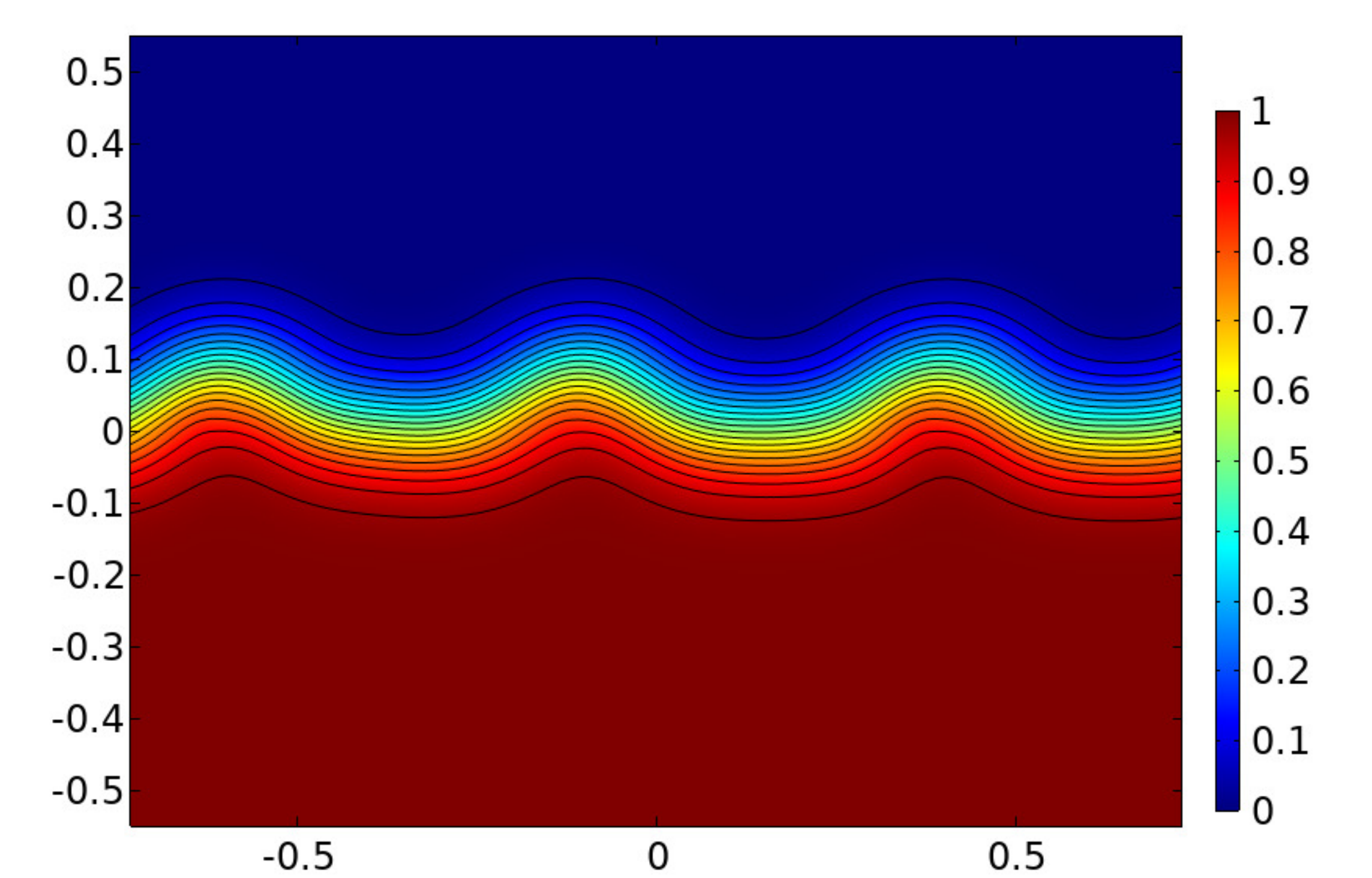}\\
  \includegraphics[width=0.8\textwidth]{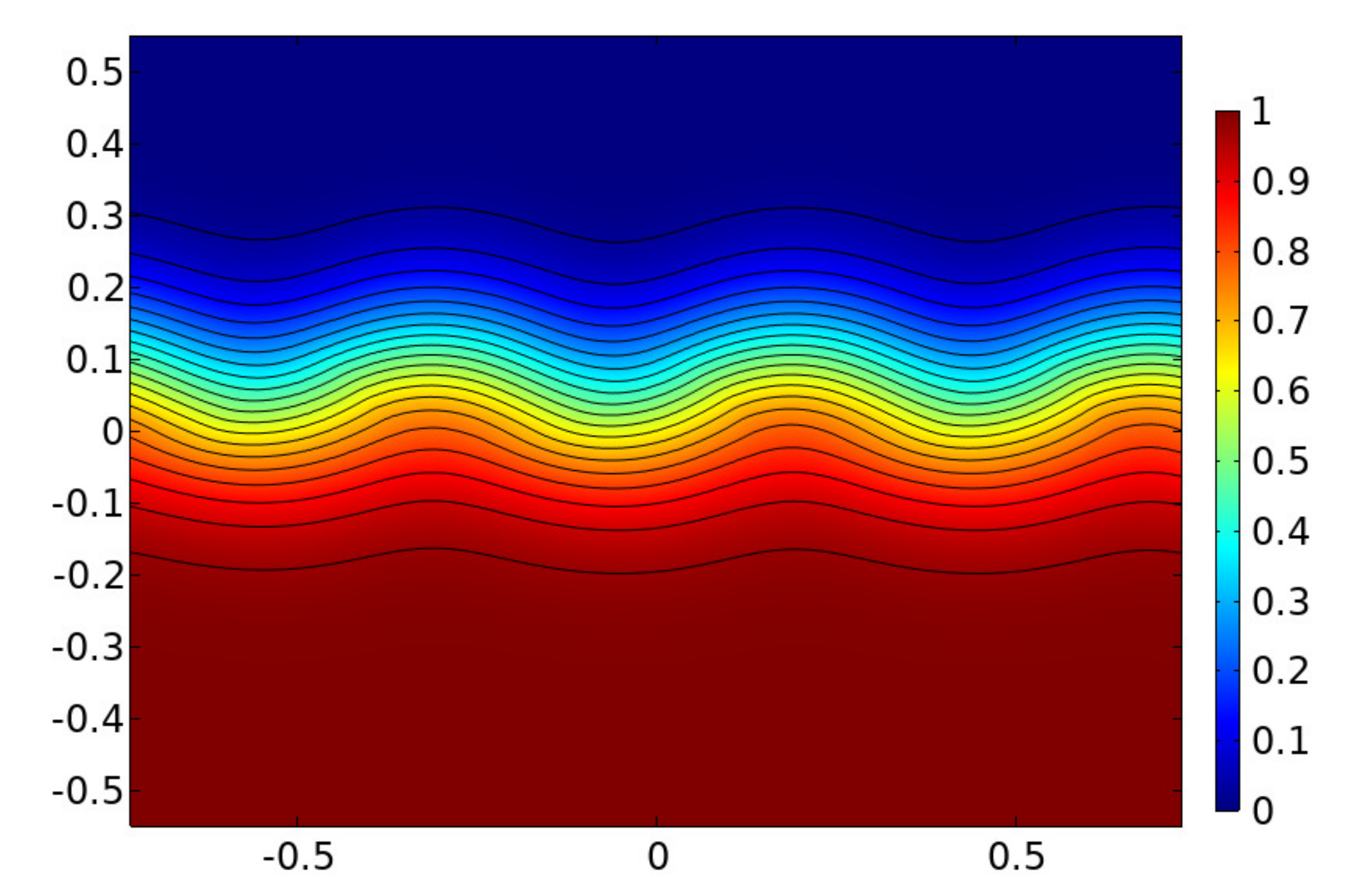}\\
  \includegraphics[width=0.8\textwidth]{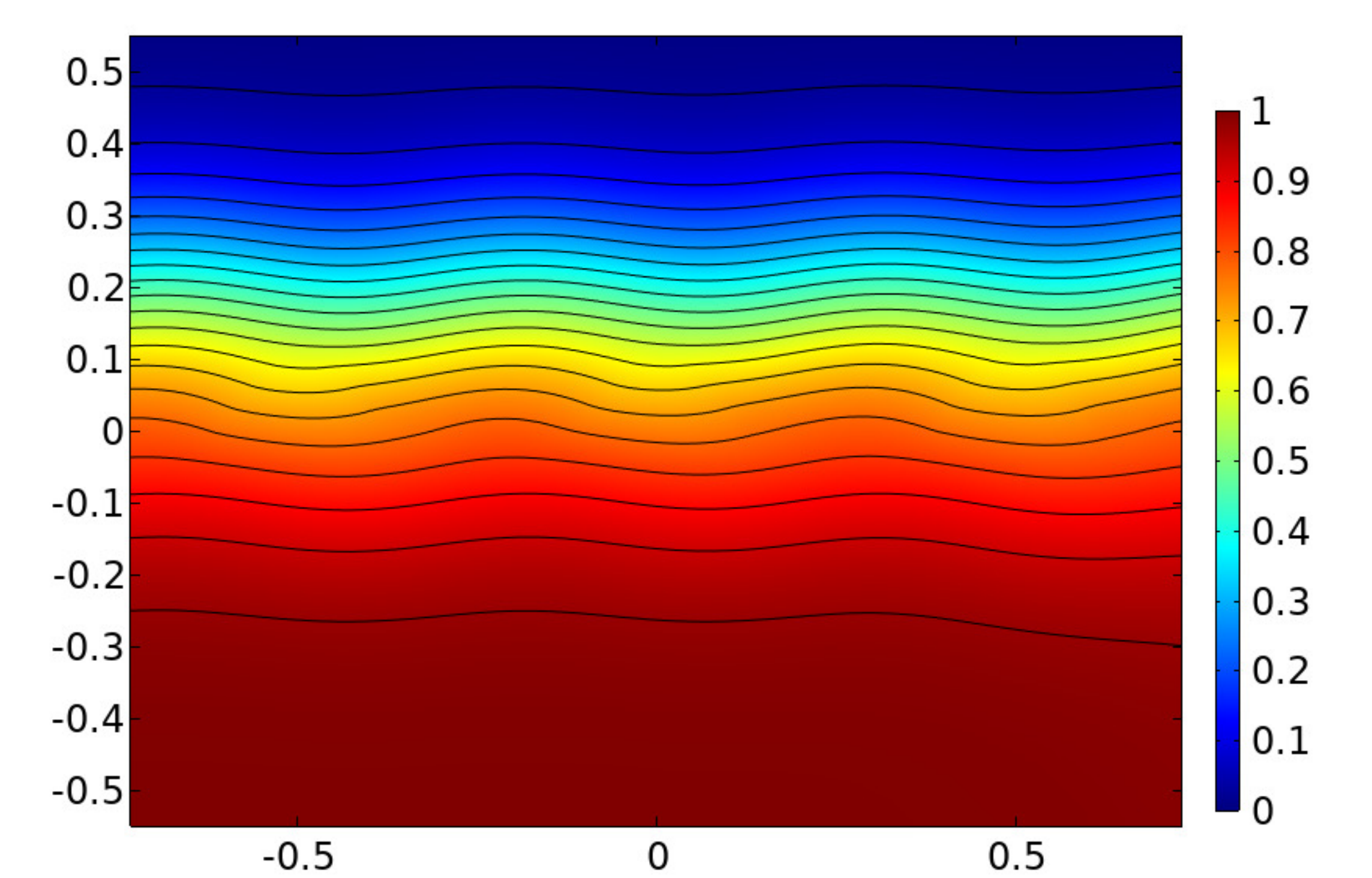}
  \caption{Electronic density levels evolution at times 0.1, 0.3 and 0.7 from top to bottom. The initial perturbation of the electronic density moves towards left, drifted by the electric field while decreases its amplitude.\label{lambda025}}
  \end{center}
\end{figure}

\begin{figure}
  \begin{center}
  \includegraphics[width=0.8\textwidth]{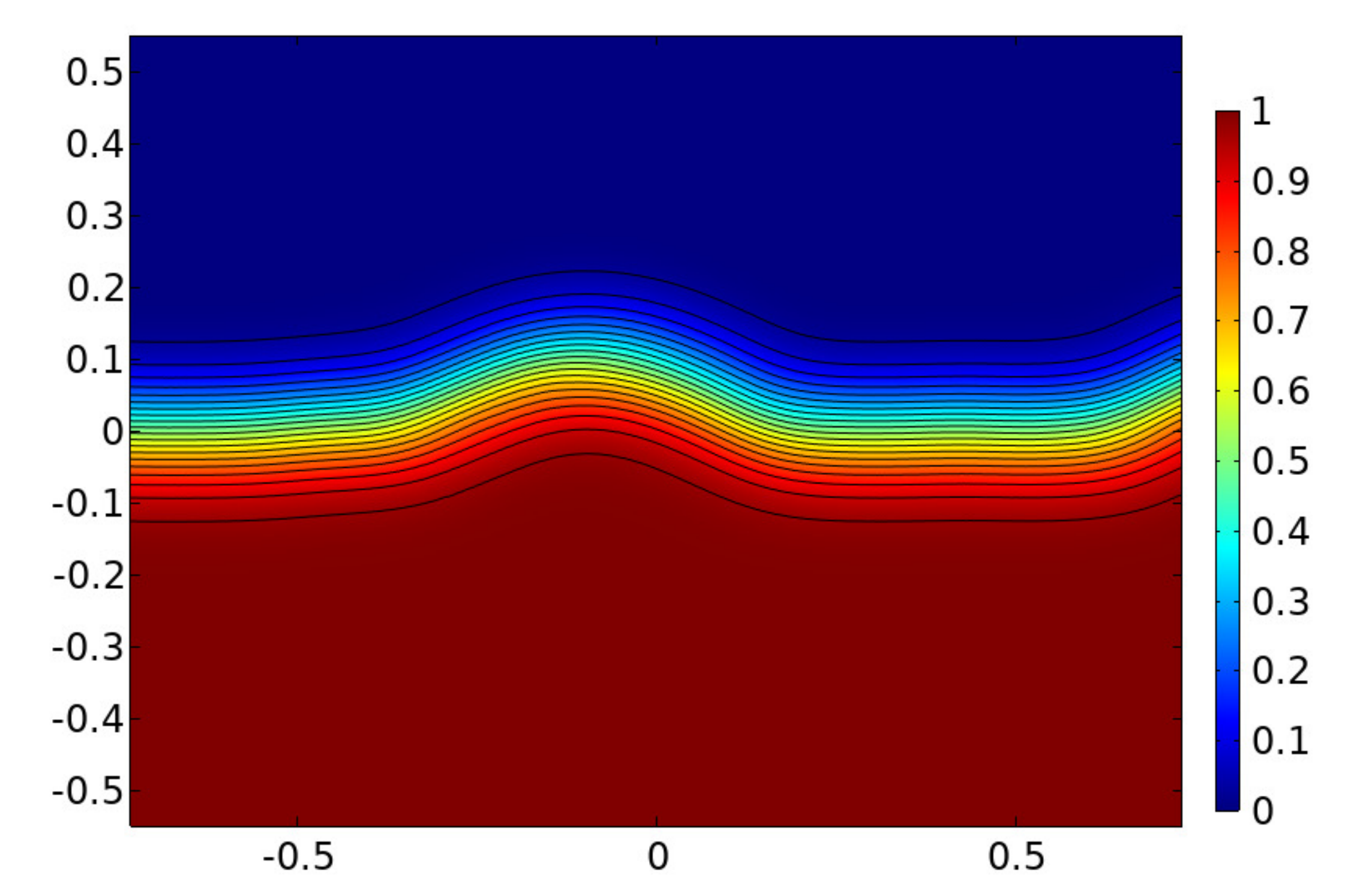}\\
  \includegraphics[width=0.8\textwidth]{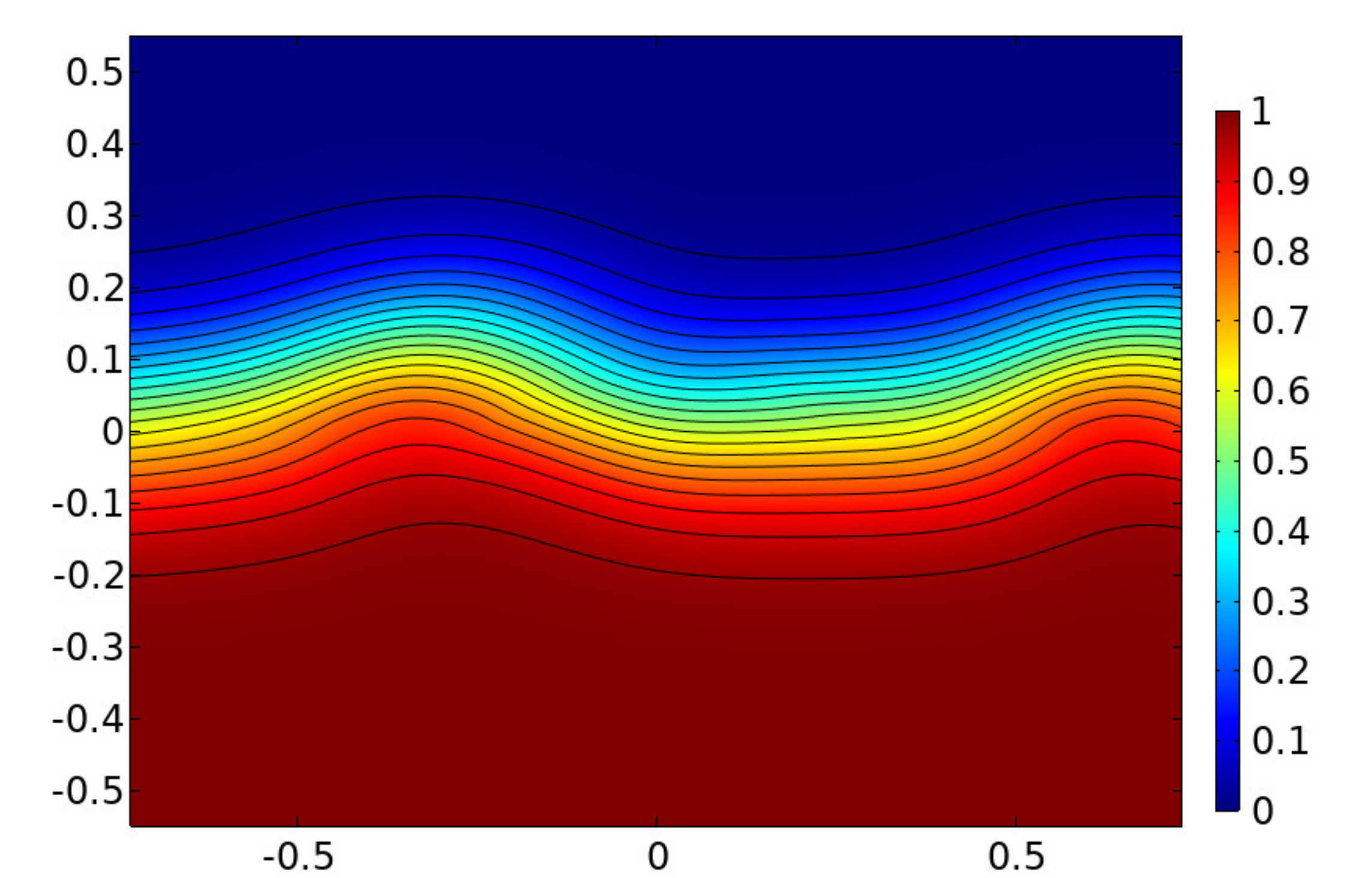}\\
  \includegraphics[width=0.8\textwidth]{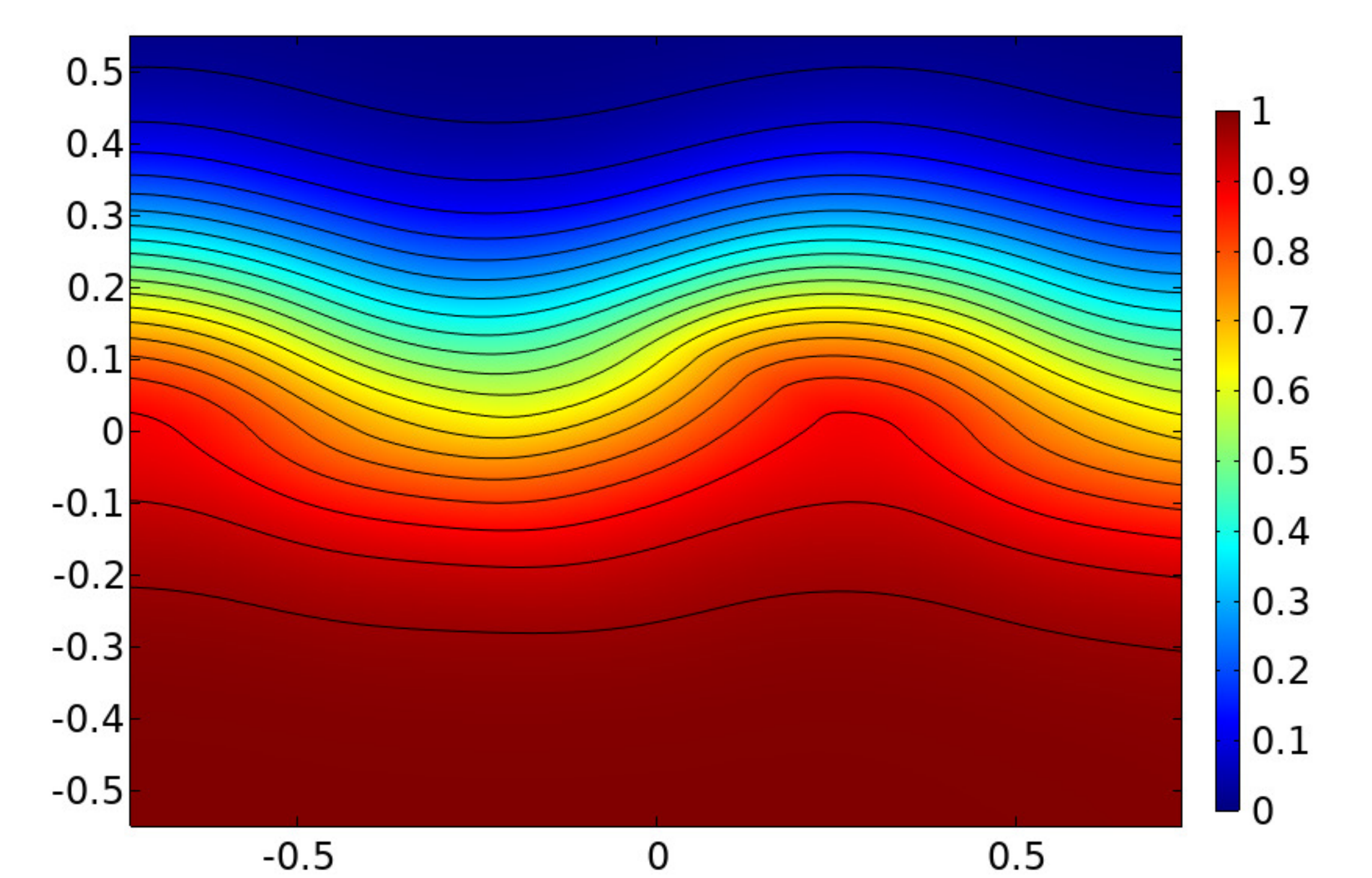}
  \caption{Electronic density levels evolution at times 0.1, 0.3 and 0.7 from top to bottom. The initial perturbation of the electronic density moves towards left, drifted by the electric field but the amplitude increases.\label{lambda1}}
  \end{center}
\end{figure}

 The numerical simulations in Fig.~\ref{lambda025}-\ref{lambda1} correspond to take the parameter appearing in the system \eqref{gauss} $\alpha=1$ and $D=0.02$ for the diffusion coefficient. To study the stability of the channel, we introduce a geometrical perturbation of the interface adding a sinusoidal displacement to the densities distributions around $z=0$ and then following the evolution. 
 
The simulations presented in Fig.~\ref{lambda025}-\ref{lambda1} has
been realized in a bigger box of $2\times 2$ of dimensionless units
using a finite element method. We used the commercial package COMSOL
Multiphysics software to perform the FEM modeling \cite{COMSOL}. In the figures we plot the electron densities at different times 0.1, 0.3, 0.7.  

As expected we observe that the perturbation moves to the left, driven by the electric field. At the same time the ionized region expands, and the channel increases the size in the $z$-direction due to creation of charge by impact ionization and diffusion. We plot two cases, one with a small wave length and other with a bigger wave length. According with the previous analysis, perturbations of wave lengths smaller than a critical value will disappear while bigger than the critical value will growth.

In Fig.~\ref{lambda025} we have introduced a small wave
length perturbation of 0.25 in dimensionless unit. We observe that the
perturbation diminishes and eventually disappears. However, when the wave length perturbation is 1, the perturbation increases as shown in Fig.~\ref{lambda1}
 
\section{Conclusion}
Using a basic model of a discharge channel we have been able to identify an instability mechanism. We have found evidence of the onset of long wave instabilities, similar to the Kelvin-Helmholtz type in fluid dynamics. The evidence of the growth of long wave perturbations has been obtained both theoretically and numerically, by asymptotic analysis of the geometrical perturbation of a planar travelling wave and solving the model using a finite element method. We note that the development of lateral instabilities creates inhomogeneities of the electric field in a similar fashion as Kelvin-Helmholtz instability in fluids creates perturbations in the otherwise one-directional velocity field. 

Let us remark that additional effects could be incorporated into the model, such as as photoionization, magnetic fields, interactions between two or more channels, etc. Each new physical effect incorporates an extra characteristic length to the problem and elucidate the dominant instability mechanism could be the next step to address the robustness of the instability mechanism that we have described here. 

\section*{Acknowledgment}
M. Array\'as and M.A. Fontelos are supported by the research Grants from the Spanish Ministry of Economy and Competitiveness, ESP2017-86263-C4-3-R and MTM2017-89423-P respectively.

\bibliography{bibfilecom}

\end{document}